\documentclass[aps,twocolumn,groupedaddress,showpacs]{revtex4}
\usepackage{epsfig}
\usepackage{graphicx}

\newcommand{\ba}{\begin{eqnarray}}
\newcommand{\ea}{\end{eqnarray}}
\newcommand{\beq}{\begin{equation}}
\newcommand{\eeq}{\end{equation}}
\newcommand{\g}{\gamma}
\newcommand{\apjl}{ Astrophys.\ J.\ Letters}
\newcommand{\aap}{ Astron.\ Astrophys.}

\begin{document}
\bibliographystyle{apsrev}

\title{Gamma Rays from the Tycho Supernova Remnant: Leptonic or Hadronic\,?}

\author{Armen Atoyan$^1$}\author{Charles D. Dermer$^2$}
\affiliation{$^1$Department of Mathematics, Concordia University,
1455 de Maisonneuve Blvd.\ West,\\ Montreal, Quebec H3G 1M8, Canada;
 atoyan@mathstat.concordia.ca\\
$^2$Code 7653, Naval Research Laboratory, Washington, DC 20375-5352; charles.dermer@nrl.navy.mil}

\date{\today}

\begin{abstract}
Recent {\it Fermi} and {\it VERITAS}  observations of the prototypical Type Ia supernova
remnant (SNR) Tycho have discovered $\gamma$ rays with energies $E$ in the range
0.4~GeV $\lesssim E \lesssim  10$ TeV.
 Crucial for the theory of Galactic cosmic-ray origin is whether the $\gamma$ rays
from SNRs  are produced by accelerated hadrons (protons and ions), or by relativistic electrons.
Here we show that the broadband radiation spectrum of Tycho can be explained within the
framework of a  two-zone leptonic model, which is likely to apply to every SNR.
A model with hadrons can also fit the radiation spectrum. The hadronic origin of $\gamma$-rays
can be confirmed  by {\it Fermi}\, spectral measurements of Tycho and other SNRs  at  $\lesssim 300$\,MeV.

\end{abstract}


\maketitle

\section{Introduction}

 Tycho (G120+1.4, 3C 10)  is the bright spherical supernova remnant
(SNR) of SN 1572  \cite{kle79} with a diameter of  $\approx 8^\prime$.
The distance $d$ to the source is not well-established, but is in the range
$d \approx 1.5$ -- 4 kpc (e.g., \citep{smi91}).
The narrow $\theta_{\rm rim} \approx 4^{\prime\prime}$ spherical rim of
hard X-ray emission is coincident with the outer edge of the radio emission
\citep{hwa02} and is explained by synchrotron emission
of relativistic electrons accelerated
 to TeV energies \citep{war05} by a forward shock moving with speed
 $v_{\rm sh} \approx 4600 \,(d/2.3\,\rm kpc) \, \, km \, s^{-1}$ \citep{Hug00}.
Infrared observations from AKARI \citep{ish10} between 9 and 160 $\mu$ reveal thermal
dust emission  in the NE and NW shells. The soft X-rays also reveal a rich
line structure implying significant thermal radiation \citep{bad06}.

Gamma-ray fluxes from Tycho between $\approx 1$ and 10 TeV have been recently reported by the VERITAS
collaboration \citep{acc11}.
The Fermi LAT Collaboration announced a detection in the energy range from
$\approx 0.4$ to 60~GeV \citep{gio11}, with energy spectral  index $\alpha \simeq 1.3$.
The total Fermi and VERITAS spectra are described by a single power law with
$\alpha \approx 1.1\! - \! 1.2$ from $\approx 500$ MeV to 10 TeV.
Nonthermal emission at  lower frequencies, from radio through X-rays,
 is universally attributed to synchrotron
radiation, which is a signiture of primary
electron acceleration.

In leptonic models, the $\gamma$ rays could result from electron
bremsstrahlung and/or  Compton-scattering processes. In hadronic models the $\gamma$ rays
arise from the decay of, predominantly,  $\pi^0$
mesons produced in interactions of cosmic-ray protons with gas and dust in the remnant.
By modeling its broadband radiation fluxes, \citet{mc11} argue that the reported $\gamma$-ray fluxes
provide strong evidence for hadronic acceleration in Tycho as the leptonic models seem to fail.
A similar conclusion was derived by
\citet{vbk08} on the basis of the then-existing upper limits of the GeV-TeV
$\gamma$-rays.
Because of the fundamental importance to the theory of SNR origin of
Galactic cosmic rays, here we undertake a detailed analysis of the robustness of those claims.

Implicit in the previous modeling is that the detected synchrotron and Compton emissions are
both produced by the very same population of relativistic electrons. This is the working
assumption of the commonly used single-zone model approach. In a SNR
environment with spatially non-uniform magnetic fields a more realistic description
of the nonthermal emission should allow at least two zones  \citep{ato00a,ato00b},
where zone 1 represents a strong magnetic-field region in the shock vicinity
where acceleration is strong, and zone 2 adjacent (further downstream ) to zone 1,
with larger volume but lower magnetic field, into which the zone 1 electrons escape
 by convection with plasma and/or by diffusion.
Because the synchrotron emissivity in zone 2 is significantly lower than in zone 1, while the
 target photon field in
both zones is basically the same, the Compton $\gamma$-ray fluxes in the two-zone model can
greatly exceed the predictions of the single-zone model.

Within this framework, we show that a leptonic model alone can explain well the
reported radio through $\gamma$-ray data of Tycho. The  TeV $\g$ rays are made primarily
by Compton emission, and bremsstrahlung produces most of the radiation at GeV energies.
The two-zone model is described in Section 2, model parameters and
interpretation of the Tycho's broadband spectrum are given in Section 3.
As concluded in Section 4, the most important confirmation of the hadronic cosmic ray
hypothesis remains the detection of the $\pi^0$-decay feature.

\section{Two-Zone Model for SNRs}

It is widely believed that cosmic rays are accelerated through Fermi processes by SNR shocks.
Both first-order (shock) and  second-order (stochastic) processes rely on differences
of magnetic field in different regions over which particle diffuse.
Strong magnetic fields favored for efficient acceleration can be produced in the vicinity
of the strong forward shock through non-linear amplification of the magnetic turbulence/Alfven
waves by the cosmic-ray ions accelerated at the shock (e.g., \citep{bl01,bell04}).

The narrow $\theta_{\rm rim} \approx 4^{\prime \prime}$ spherical rim of hard X-ray
emission \citep{hwa02} is evidence for such processes operating at the forward shock of Tycho.
Interpretation of the rim width $h\approx 1.93\, 10^{-2} d_{\rm kpc} \, \rm pc$, at
the source distance $d_{\rm kpc}\equiv d/1 \, \rm kpc$, as due to fast synchrotron cooling of
$\gtrsim 10 \,\rm TeV$ electrons  implies a high
magnetic field in the rim, $B\gtrsim 400 \,\rm \mu G$ \citep{vbk08}. In effect, this explanation leads to
a single-zone model where all the TeV $\gamma$-rays and the X-rays are assumed
to be produced by the
same electrons.

However, the rim could also be explained as the region with strongly enhanced magnetic
field behind the shock. The rim width then reflects the length scale of damping the
magnetic turbulence  rather than cooling of TeV electrons \citep{pyl05} (see also the
discussion in \citep{rgb11}).
Relativistic electrons accelerated at the rim escape from
zone 1 (the rim) into the zone 2 by diffusion and/or convection with the fluid.
The theory of cosmic-ray transport is highly developed, and involves at least
the spectrum and anisotropic character of the turbulence \citep[e.g.,][]{gs97,mlp06}.
Nevertheless, reasonably accurate general estimates can be made. For Tycho the convective
escape time $\tau_{\rm c} =h/v_{\rm fl}^\prime \approx 40 \, \rm yr$, where
$v_{\rm fl}^\prime\approx v_{\rm sh}/4 \approx 500 \, d_{\rm kpc} \,\rm km \, s^{-1}$ is the
fluid speed in the forward shock frame. The diffusive escape time from zone 1 is $\tau_{\rm dif}
\approx h^2/(2\kappa)$ where $\kappa = \lambda_{\rm sc} c/3$ is the diffusion
coefficient. The mean scattering length $\lambda_{\rm sc}$  of relativistic particles
is equal to the gyroradius $r_{\rm gyr}$ in the Bohm diffusion limit,
which is attained if the ratio of the turbulent magnetic field fluctuations to the
mean field is $\eta = |\delta B|/B \simeq 1$. A value $\eta < 1$ is probably more realistic in
general. In this case
$\lambda_{\rm sc}\approx r_{\rm gyr}/\eta^2$ (e.g., \citep{bell04}).

The diffusive escape time
of electrons with Lorenz factor $\gamma$ from the rim  can be estimated as
$\tau_{\rm dif} \approx 1.1 \times 10^4 \, \eta^2 \, B_{\rm \mu G}
d_{\rm kpc}^2 \, \gamma^{-1} \, \rm yr$. Comparing this with the synchrotron cooling
time $t_{\rm syn} \approx 2.45\times 10^{13}\, B_{\rm \mu G}^{-2} \gamma^{-1}\,$yr, one
finds that accelerated electrons (of any energy) could cool in the rim before
escaping downstream only if $B \geq 1.31 \,\eta^{-2/3} \rm mG$.  Even in the limit
of $\eta = 1$ this field is high. For any smaller magnetic
field, electron escape from the acceleration region
before cooling cannot be prevented. We note that this very effect explains the detection of
non-thermal X-rays  not only from the rim but also from a wider
 region interior to the rim.

Weaker magnetic fields interior to the thin rim could be the result of
  dissipation of the magnetic turbulence in the thermal plasma downstream of the forward shock
on timescales $\tau_{\rm c}$ yr.
Therefore the region of the shell interior to the rim forms  zone 2.
This zone extends at least down to the contact discontinuity at
$r_{\rm CD}/r_{\rm sh}\approx 0.927$ \citep{war05}, in which case the volume
$V_2$ of zone 2 is larger by a factor $\approx 3.3$ than the zone 1 volume $V_1\approx 0.05\,V_{\rm SNR} $.
In a more elaborate model, the region between the contact
discontinuity and the reverse shock at $\theta \approx 183^{\prime \prime}$
would be interpreted as another zone with a different set of source parameters. However, because
the prime goal of this paper is to establish the need for  multi-zone
modeling, which is appropriate
for spatially inhomogeneous sources and can  significantly relax the constraints
on the synchrotron and Compton fluxes compared to the single-zone model, here we limit our
calculations within the framework of two zones only. This assumes that
zone 2 includes most of the shell between the thin X-ray rim (the zone 1) and the reverse shock.
This is qualitatively justified as the
contact discontinuity is apparently porous, with
both thermal gas and relativistic particles penetrating through the contact
discontinuity further down, e.g., through Rayleigh-Taylor instabilities as
evident from detection of thermal X-ray filaments
\citep[e.g.,][]{hwa02,bad06}.
The implied volume filling factors of the two zones in the remnant are
$\zeta_1=V_1/V_{\rm SNR}\approx 0.05$ and $\zeta_2=V_2/V_{\rm SNR}\lesssim 0.5$.
These values can accomodate both interpretations of the X-ray stripes in Tycho
\cite{2011ApJ...728L..28E} as part of either zone 1 (as further inhomogeneities in the
rim \cite{2011ApJ...728L..28E}) or zone 2 (inhomogeneities in the shell).

The system of coupled equations describing the energy distribution functions
$N_1(E,t)$ and $N_2(E,t)$ of electrons (or protons)
in a two-zone model was derived in \citep{ato00a}:
\begin{equation}
\frac{\partial N_1}{\partial t} = \frac{\partial (P_1\,N_1 )}
{\partial E}  - \frac{N_1}{\tau_{\rm c}} - \left( \frac{N_1}{V_1}
 -\frac{N_2}{V_2}\right)\! \frac{V_1}{\tau_{\rm dif}}
 +\, Q_1
\end{equation}
\begin{equation}
\frac{\partial N_2}{\partial t} = \frac{\partial ( P_2\,N_2 )}
{\partial E} + \frac{N_1}{\tau_{\rm c}} + \left( \frac{N_1}{V_1}
 -\frac{N_2}{V_2}\right)\! \frac{V_1}{\tau_{\rm dif}}
 +\, Q_2
\end{equation}
Here $P_{1} \! \equiv \! P_1 (E,t)$ and $P_2 \! \equiv \! P_2 (E,t)$ represent the energy-loss rates
$(-d E/ d t)$ of particles with energy $E$ in zones 1 and 2 respectively. The second term in the right
side describes convective escape of electrons from zone 1 into zone 2, and the 3$^{rd}$  term describes
the diffusive escape of electrons from the zone 1 into zone 2 and vice versa proportional
to the {\it difference} of the spatial densities ({\it the gradient}) of their energy
distributions in the transition region between the zones. This leads to vanishing of the
diffusive exchange of particles between the zones at very high energies when
the energy densities in the two zones become equal. The timescale
$\tau_{\rm dif}(E)$ depends primarily on the diffusion rate in zone 1 because diffusion their
is much slower (see \citep{ato00a} for details).
$Q_1\equiv Q_1(E,t)$ and $Q_2 \equiv Q_2(E,t)$ are the acceleration rates
of particles in zones 1 and 2, respectively.


The integral form of the solution for this type of equationsegions of
 can be used to find  a self-consistent solution for the
system using the numerical method of iterations \citep{ato00b}, neglecting  the diffusive
influx of particles from zone 2 to zone 1 in the first iteration.
Calculations show that $\lesssim 50$ iterations result in a
stable (converged) solution for $N_1(E,t)$ and $N_2(E,t)$.

\section{Model Parameters and Radiation Fluxes for Tycho SNR}

Fluxes of $\gamma$-rays produced by relativistic
electron bremsstrahlung or by hadronic $pp$-interactions are proportional to the
target gas density $n_{\rm p}$ (in terms of hydrogen atoms) in the source.
The gas density $n_1$ in zone 1 (the rim) is proportional to the ambient gas density $n_0$ upstream
of the forward shock, and is about $n_1 \approx 4 \, n_0$ for a strong shock.
Modeling of the X-ray emission lines suggests \citep{bad06} the best fit ambient gas density
$\rho_0=m_{\rm p} n_0 \simeq 2 \times 10^{-24} \, \rm g\, cm^{-3}$,
or $n_0\approx 1.2 \,\rm cm^{-3}$ (see also \cite{hss86}).  Smaller values, in the range
$n_0 \simeq (0.3- 1)\,\rm cm^{-3}$ for distances $d \simeq (3-2.35) \,\rm kpc$, respectively,
have been inferred from hydrodynamic modeling of the
X-ray emission of forward-shocked material \citep{cas07}.

 The mean gas density $n_2$ in zone 2 (the shell) can be derived from the estimate of the total
gas mass accumulated in zone 2, $M_{2}\! \approx \! m_p n_2 \zeta_2 \, V_{\rm SNR}$, where
$\zeta_2$ is the volume filling factor. Note that in general this mass is contributed not only
from $n_0$, but also from the mass of the pre-supernova wind swept up by the forward
shock, and the mass $M_0$ of the SN ejecta.
 To calculate $M_2$  we note that the measured speed of the forward shock
is significantly smaller than the initial
$v_0\approx 10^4\, (E_{\rm SN.51})^{1/2} \, M_{0.\odot}^{-1/2}\, \rm km \, s^{-1}$.
For a typical Type Ia SN ejecta with ejecta mass $M_{0.\odot} = (M_0/M_\odot) \approx 1 $
and kinetic energy $E_{\rm SN.51}=E_{\rm SN}/ 10^{51} \rm erg \approx 1 $ this implies
up to several Solar masses residing in the shell of Tycho.

Indeed, most of the initial explosion energy should still be in kinetic
form as both the total irradiated energy and the energy accumulated in accelerated
particles are much smaller than $E_{\rm SN}$. The total energy of the thermal
electrons with temperature $k T_{\rm e} \sim 1\,\rm keV$ deduced from X-ray
observations \citep{hwa02, war05, bad06} is only
$E_{\rm T,e} = (3k T_{\rm e}/2) M_{2}/m_{\rm p}
\approx 2.5 \times 10^{48} (M_2/M_\odot) \,\rm erg$. To contribute substantially to the
total energy budget in the shell, thermal proton plasma should have a temperature
 of order $T_{\rm p} \gtrsim 10^{9} \,\rm K$. This is unlikely given fast heating of the
 electrons through Coulomb interactions with protons in such two-temperature plasma.
Thus, the kinetic energy of the fluid moving
 with $v_{\rm f} \simeq 3v_{\rm sh}/4$ behind the forward shock in the shell
  should be still close to the initial energy,
 $E_{\rm kin} = M_{2} v_{\rm f}^2/2 \simeq 10^{51}E_{51}\, \rm erg$.
 For $v_{\rm sh}$ $\approx 4600 \, (d/2.3) \,\rm kpc$ \citep{Hug00} the implied gas density
 in zone 2
 \begin{equation}
 n_2 \approx 1.05 \,\, \zeta_2^{-1} \, (d /3 \, {\rm kpc})^{-5} E_{51}
  \; \rm cm^{-3} \, .
 \end{equation}
For $\zeta_2 = 0.5$ this leads to $n_2$ in the range
from $ 5 \, \rm cm^{-3}$ for $d = 2.5\,\rm kpc$ to $ \approx 0.9 \, \rm cm^{-3} $ for $d = 3.5 \,\rm kpc$.
In calculations here we use $n_2=3\,\rm cm^{-3}$ corresponding to $d\approx 2.8\,\rm kpc$.

For the Compton radiation we take into account not only the
CMB target photons, but also the FIR photons produced by thermal dust in the NE and
NW parts of Tycho \citep{ish10}. This component can contribute
up to 30\% to the total Compton flux at TeV energies.

The particle injection rate $Q_1$ is in general time-dependent, reaching the maximum
at the transition time to Sedov phase, and gradually declining afterwards. However, taking into
account that Tycho is still in its early Sedov phase,
we assume a stationary injection rate
$ Q_1(E,t)\propto \, E^{-\alpha} e^{-E/E_{\rm cut}}$
with $\alpha=2.3$ and $E_{\rm cut}=36 \,\rm TeV$.
Given that the reverse shock in Tycho is not a bright non-thermal
X-ray source, injection of electrons in zone 2 at the reverse shock is presumably much
smaller than at the forward shock. To keep the model simple,
we neglect electron acceleration in zone 2, i.e. $Q_2 = 0$.

\begin{figure}
{\includegraphics[width=\columnwidth]{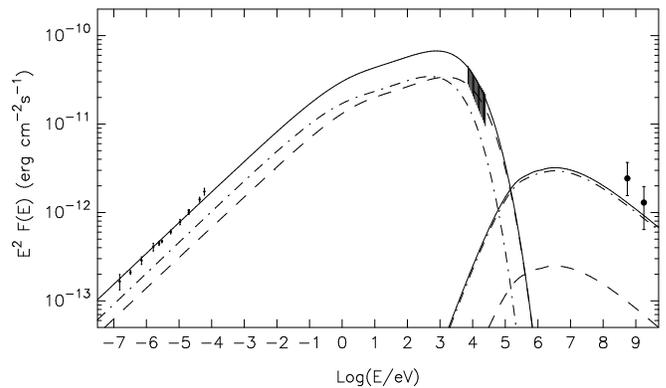}}
\caption[]{
The synchrotron fluxes from radio through X-rays
in the two-zone model. Dashed and dot-dashed lines show the fluxes from zone 1 and
zone 2, respectively, and the total flux is shown by the solid line.
Calculations assume density $n_2\approx 3 \,\rm cm^{-3}$ at $d_{\rm kpc} =2.8$, and $n_1 \approx n_2$.
Other model parameters are $B_1= 100 \,\rm \mu G$
and $B_2= 32 \,\rm \mu G$, $\eta=0.3$,
$\alpha = 2.3$ and $E_{\rm cut}=36 \,\rm TeV$. Also shown are
$\lesssim\,$GeV bremsstrahlung fluxes produced by
relativistic electrons in zones 1 and 2.  }
 \label{Fig1}
\end{figure}

\begin{figure}
\includegraphics[width=\columnwidth]{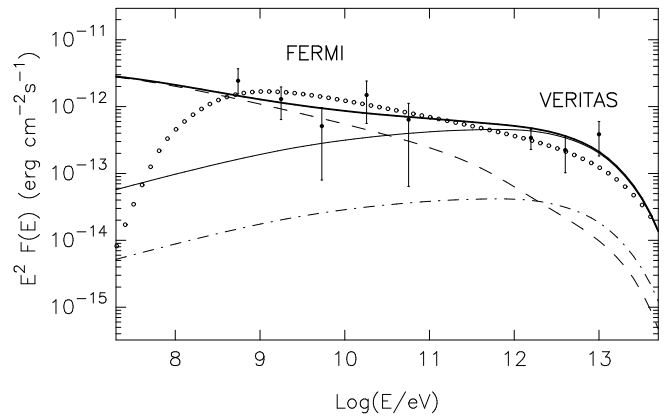}
\caption{
The $\gamma$-ray fluxes produced in the two-zone model.
The heavy solid line shows the total flux of leptonic origin.
The total bremsstrahlung and Compton radiation fluxes are shown by dashed and solid
(thin) lines, respectively. Both are predominantly produced in zone 2.
For comparison, the Compton flux contribution from zone 1 is also shown (dot-dashed line).
The full open dots show a possible fit to data by a hadronic model (see text).
}
\label{Fig2}
\end{figure}

Figure 1 shows the synchrotron fluxes from radio through X-rays
calculated assuming $B_1=100\,\rm \mu G$ and $B_2 = 32 \,\rm \mu G$ in zone 1 and
zone 2, respectively.
In the radio to sub-mm wavelengths
the total flux is contributed mostly by zone 2, while in non-thermal X-rays
zone 1 contributes more than zone 2. This picture is in qualitative agreement
with observations that the narrow rim is much brighter in X-rays than in radio.

Figure 2 shows the $\gamma$-ray
fluxes. The total flux of leptonic origin
is dominated by bremsstrahlung up to $\sim 100\, \rm GeV$, most of which
is produced in zone 2 where most of electrons accelerated in zone 1 eventually reside.
The assumed mean gas density $n_2\approx 3 \,\rm cm^{-3}$ in zone 2, at $d=2.8\,\rm kpc$, implies
about $3\,M_\odot$ of gas accumulated in the shell of Tycho. The gas density $n_1$ in zone 1 doesn't
have any significant impact on the calculated fluxes. A density $n_1$ in the rim about the same as the
average $n_2$ in the shell would imply an ambient gas density $n_0\approx n_1/4 = 0.75 \,\rm cm^{-3}$.
Even lower gas densities $n_0$ cannot, however, be excluded taking into account that the mass
currently accumulated in the shell may have a composition dominated not by the swept up
ambient gas but by the presupernova wind.

 The total energy of the electrons in zones 1 and 2 are
$E_{e.1}=4.4 \times 10^{47} \,\rm erg$ and $ E_{e.2}=4.3 \times 10^{48} \,\rm erg$,
respectively.
The thin solid line shows the total Compton flux. Because the target
photon field (CMBR + FIR) penetrates easily through the entire source and thus is same in
both zones, the Compton contribution from zone 1 is insignificant.

The detected $\gamma$-ray fluxes can also be explained by relativistic protons.
Full open dots in Figure 2 show the flux from $\pi^0$ decay $\gamma$-rays produced by protons
with the power-law index $\alpha_{\rm p}= 2.3$ and the total energy $E_{p}=3\times 10^{49} \,\rm erg$.
We emphasize that the {\rm essential} difference between the hadronic and leptonic fluxes occurs
only at energies below $\approx 500 \,\rm MeV$. Therefore only by detecting the characteristic
$\pi^0$ decay cutoff in the energy spectrum of $\gamma$-rays at $E\simeq (100-300)$ MeV will it be possible
to claim that cosmic ray hadrons are the origin of the $\gamma$-rays from Tycho.

\section{Summary and Discussion}

Clear identification of  the $\pi^0$ decay bump peaking at $E = m_{\pi^o}/2$ in a photon
spectrum was proposed long ago as a promising way to establish acceleration sites of cosmic-ray
protons and nuclei \citep{gs64,hay69}. The Large Area Telescope on {\it Fermi} provides a new
database for tackling this question, and the
Type Ia SNR Tycho presents one of the best opportunities to search for
signatures of hadronic production.

Any realistic treatment of the acceleration and radiation must consider at least two zones
of different magnetic field strength. Electrons in the thin acceleration zone behind the shock
where the field is enhanced are bright in synchrotron X-rays.
Electrons in the region with the weaker field (zone 2) occupying the bulk of the shell volume
will have dim synchrotron emission, and can be essentially missed in single-zone analyses.
This zone can, however, provide most of the Compton and bremsstrahlung flux.
This effect significantly weakens limitations inferred from a single-zone analysis of the
observed synchrotron X-ray and GeV-TeV $\gamma$-ray fluxes.
With a more realistic two-zone model, we find that a good fit to the Fermi  and VERITAS data from the
Tycho SNR can be made by a purely leptonic model, with bremsstrahlung emission making, primarily,
the GeV $\gamma$ rays,
and TeV photons coming mainly from Compton-scattered soft radiation fields.
No hadrons are required in this model.
The underlying densities and volume filling factors of the two zones
are not uniquely parameterized, our solution represents only one of a range of possibilities.

Figure 2 shows that hadronic models can fit the data just as well as a leptonic model. The two models
are quite different only at low energies, with the hadronic model exhibiting
the $\pi^0$-decay feature, shifted up to a few hundred MeV in a $\nu F_\nu$ representation (e.g. \citep{der86}).
The absence of this  this low energy feature would
rule out a hadronic origin, while its detection would support
an origin from cosmic-ray protons and ions.

Establishing the flux of Tycho
at low energies with the {\it Fermi}-LAT is difficult because of the reduction in the effective area and
the larger point spread function below 1 GeV \citep{atw09}. Moreover, background modeling
of the diffuse $\gamma$-ray emission from the Galaxy disk has to be carefully
treated. Progress in low-energy analysis and longer exposure times will
help improve the quality of the {\it Fermi}-LAT spectrum of Tycho. The kinematic
$\pi^0$ flux reduction is, however, a generic feature of cosmic-ray hadron origin from the GeV
cosmic rays that carry most of the energy.  Identification of this feature in Tycho
or other SNRs will finally provide experimental confirmation of the SNR origin of the Galactic cosmic rays.

\vskip0.1in

\noindent
We thank J.\ Ballet, M.\ Laming, J.\ Finke, S.\ Funk, S.\ Razzaque, and S.\ Reynolds for discussions.
AA appreciates the support and hospitality of the NRL Gamma and Cosmic
Ray Astrophysics Branch during his visit
 when this work was initiated. The work of C.D.D.\ is supported by the Office of Naval Research.

\end{document}